\documentclass[12pt,a4paper,final]{iopart}

\usepackage{iopams}  
\usepackage{graphicx}
\usepackage{makecell}
\usepackage[breaklinks=true,colorlinks=true,linkcolor=blue,urlcolor=blue,citecolor=blue]{hyperref}
\usepackage{pdflscape}
\usepackage{textcomp}
\usepackage{soul}
\usepackage{changepage}

\begin{document}

\title[Stopping-power ratios for carbon ion beams]{Impact of new ICRU90 key data on stopping-power ratios and beam quality correction factors for carbon ion beams}

\author{Lucas Burigo$^{1,2}$}
\address{$^1$German Cancer Research Center (DKFZ), Heidelberg, Germany}
\address{$^2$National Center for Radiation Research in Oncology (NCRO), Heidelberg Institute for Radiation Oncology (HIRO) Heidelberg, Germany}
\ead{l.burigo@dkfz-heidelberg.de}

\author[cor1]{Steffen Greilich$^{1,2}$}
\address{$^1$German Cancer Research Center (DKFZ), Heidelberg, Germany}
\address{$^2$National Center for Radiation Research in Oncology (NCRO), Heidelberg Institute for Radiation Oncology (HIRO) Heidelberg, Germany}
\ead{s.greilich@dkfz-heidelberg.de}

\begin{abstract}
The recent update of dosimetric key data by the `International Commission on Radiation Units and Measurements' impacts the computation of beam quality correction factors $k_Q$ via several changes, e.g. for the mean excitation energies, $I$, which enters the stopping power computation for water and air, the computation procedure itself, the average energy expended in the production of an ion pair in air, $W/e$, as well as chamber perturbation factors for Cobalt-60. An accurate assessment of water-to-air stopping-power ratio, $s_{\rm w,air}$, in reference conditions with new recommendation is necessary to update the dosimetry protocols for carbon ion beams. The new ICRU90 key data were considered for computation of $s_{\rm w,air}$ for carbon ion beams using Monte Carlo transport simulations for a number of reference conditions, namely monoenergetic carbon ion beams with range in water from 3 to 30\,cm and Spread-Out Bragg peaks (SOBPs) of different widths and depths in water. New recommendations for $s_{\rm w,air}$ are presented, namely 1.1247 for the reference condition of 1\,g\,cm$^{-2}$ depth for monoenergetic carbon ion beams and 1.1274 at the center of physically-optimized SOBPs. The recommendation of a constant value (1.126) represents the stopping-power ratio within a 0.3\,\% variation of $s_{\rm w,air}$ for the different reference conditions. The impact of these new $s_{\rm w,air}$ values and the updated key data on the $k_Q$ for carbon ion beams was evaluated in a second step. The changes agree very well with experimental data for the case of cylindrical ionization chambers, but larger discrepancies are observed for plate-parallel ionization chambers.

\end{abstract}

\vspace{2pc}

\section{Introduction}

Reference dosimetry for carbon ion beams relies on calibrated ionization chamber measurements and dose-to-water-based protocols. The most prominent code of practice -- IAEA's TRS-398~\cite{IAEA2000} -- defines the relation between dose-to-water $D_{\rm{w},Q}$ for this beam quality $Q=\,^{12}\rm{C}$ and the charge measured by the chamber as:
\begin{equation}
D_{\rm{w},Q} = M_Q \cdot N_{D,\rm{w},Q_0} \cdot k_{Q,Q_0}
\end{equation}
where $M_Q$ is the corrected chamber reading. Since no primary standard for carbon beams exists, the chamber calibration coefficient $N_{D,\rm{w},Q_0}$ from $Q_0=\,^{60}$Co is multiplied with the `beam quality correction factor' $k_{Q,Q_0}$. Experimental data on $k_{Q,Q_0}$ are still scarce and $k_{Q,Q_0}$ is usually computed as originally suggested for high-energy photon beams by \cite{Andreo1992} via
\begin{equation}
k_{Q,Q_0} = \frac{\left(s_{\rm{w,air}}\right)_Q} {\left(s_{\rm{w,air}}\right)_{Q_0}} \cdot \frac{p_Q}{p_{Q_0}}\cdot\frac{\left(W/e\right)_Q}{\left(W/e\right)_{Q_0}} 
\label{eq:kq}
\end{equation}
i.e. as the ratio between the stopping-power-ratios (SPR) $s_{\rm{w,air}}$, the chamber perturbation factors $p$ and the mean energies required to produce an ion-pair in air, $W/e$, for the respective beam qualities. Radiation transport simulations with detailed chamber geometries can yield better accuracy by determining the combined effect of stopping power ratio and chamber perturbation factor, $f=s_{\rm{w,air}}\cdot p$, and should therefore be preferred. Such studies have been published for protons \cite{Goma2016} and high-energy photons \cite{Wulff2008,Muir2010} but not for carbon ions yet. 

The recent update of dosimetric key data by the `International Commission on Radiation Units and Measurements' in their Report No.~90~\cite{ICRU90} impacts Eq.~\ref{eq:kq} via changes of the mean excitation energies $I$ which enters the stopping power computation for water and air, the computational procedure of stopping power itself, and $W/e$ (Tab.~\ref{tab:quantities}). All available studies on carbon ion SPR \cite{Geithner2006,Henkner2009,Luehr2011,Goma2013,Sanchez2013} were published before ICRU90. Only Andreo \emph{et al.} \cite{Andreo2013} estimated the effect to be -0.5\,\%, mainly due to the changes in $I$-values. 

In this study, we therefore evaluated the impact of the updated key quantities on the stopping-power ratio $s_{\rm{w,air}}$ and the $k_Q$ factors for carbon ion beams in detail. We investigated both pristine and spread-out Bragg peaks (SOBP) to cover a wide range of conditions in reference dosimetry and parametrized the SPR as a function of residual range as a beam quality specifier.

\begin{landscape}
\Table{\label{tab:quantities}Recommended values for quantities relevant for the calculation of $k_{Q,Q_0}$ in the original version of IAEA's CoP (`TRS398', 2000), the ICRU73 report (2005) and its corrigendum (2009), the draft of the German regulation (`DIN6801-1'), and the new ICRU90 report (2014). Changes in ICRU90 are in bold face.}

\br
Quantity & TRS-398 & ICRU73 & DIN6801-1 & ICRU90\\
\mr
\multicolumn{5}{l}{\textit{$^{60}$Co}}\\
\mr
$s_{\rm{w,air}}$ & 1.133$\pm$0.5\,\% & --- & 1.133$\pm$0.1\,\% & \textbf{(1.127)}$^{\rm a}$ \\
$p$ & $\pm$0.6\,\% (cyl.), $\pm$1.5\,\% (pp) & --- & Partly updated$^{\rm b}$, $\pm$0.6\,\% (cyl.), $\pm$1.1\,\% (pp) & \textbf{+1.2\,\%} \\
$W_{\rm{air}}$/e & 33.97\,eV$\pm$0.2\,\% & --- & 33.97\,eV$\pm$0.2\,\% & 33.97\,eV$\pm$\textbf{0.12\,eV (0.35~\%)}\\
\mr
\multicolumn{5}{l}{\textit{Protons}}\\
\mr
$s_{\rm{w,air}}$ & Analytical expression$^{\rm c}$, $\pm$1\,\% & --- & Analytical expression$^{\rm c}$, $\pm$1.5\,\% & ---\\
$p$ & $\pm$0.8\,\% & --- & --- & --- \\
$W_{\rm{air}}$/e & 34.23\,eV$\pm$0.4\,\%$^{\rm e}$ & --- & 34.23\,e$\pm$0.4\,\%$^{\rm e}$ & \textbf{34.44\,eV}$\pm$0.14\,eV (0.4~\%)$^{\rm e}$\\
$I_{\rm{water}}$ & 75.0$\pm$2.0\,eV$^{\rm g}$ & --- & 75.0$\pm$2.0\,eV$^{\rm g}$ & \textbf{78.0\,eV}$\pm$2.0\,eV\\
$I_{\rm{air}}$ & 85.7$\pm$1.7\,eV$^{\rm g}$ & --- & 85.7$\pm$1.7\,eV$^{\rm g}$ &  85.7\,eV$\pm$\textbf{1.2\,eV}\\
Other & --- & --- & --- & \multirowcell{2}[l]{\textbf{`Improved calculation}\\ \textbf{of $S_{\rm{el}}/\rho$'}}\\
\mr
\multicolumn{5}{l}{\textit{Light ions (He-Ar)}}\\
\mr
$s_{\rm{w,air}}$ & 1.130$\pm$2\,\% & --- & Analytical expression$^{\rm d}$, $\pm$1.5\,\% & ---\\
$p$ & 1.0$\pm$1.0\,\% & & 1.0$\pm$0.1\,\% & 1.0$\pm$1.0\,\%\\
$W_{\rm{air}}$/e & 34.50~eV$\pm$1.5\,\%$^{\rm e}$ & --- & 34.50~eV$\pm$1.5\,\%$^{\rm e}$ & \textbf{34.71~eV$\pm$0.52~eV (1.5~\%)}$^{\rm f}$\\
$I_{\rm{water}}$ & 75.0$\pm$2.0 eV$^{\rm g,i}$ & 67.2\,(corr. 78)\,eV ($Z>2$)$^{\rm h}$& \multirowcell{2}[*]{75.0\,eV ($Z=2$) \\ 78.0\,eV ($Z>2$)} & \textbf{78.0\,eV}$\pm$2.0\,eV \\
$I_{\rm{air}}$ & 85.7$\pm$1.7\,eV$^{\rm g,j}$ &82.8\,eV ($Z>2$)$^{\rm h}$& \multirowcell{2}[*]{85.7\,eV ($Z=2$) \\ 82.8\,eV ($Z>2$)} & 85.7\,eV$\pm$\textbf{1.2\,eV} \\
Other & --- & --- & --- & \multirowcell{2}[l]{\textbf{`Improved calculation}\\ \textbf{of $S_{\rm{el}}/\rho$'}}\\
\br
\end{tabular}
\item[] $^{\rm a}$ Not explicitly stated (but $s_{\rm{w,air}}\cdot p_{\rm{CH}}$), isolated value given in \cite{Goma2016}.
\item[] $^{\rm b}$ From Muir and Rogers~\cite{Muir2010} -- otherwise taken from TRS398.
\item[] $^{\rm c}$ $a + b\cdot R_{\rm{res}} + c/R_{\rm{res}}$, with $a=1.137$, $b=-4.3\cdot 10^{-5}$, and $c=1.84\cdot 10^{-3}$
\item[] $^{\rm d}$ Same as $^{\rm c}$, with $a=1.130$, $b=-9.0\cdot 10^{-5}$, and $c=8.889\cdot 10^{-4}$ for alpha particles and $a=1.1203$, $b=-3.998\cdot 10^{-5}$, and $c=3.942\cdot 10^{-4}$ for $Z>2$
\item[] $^{\rm e}$ Independent of particle energy / type.
\item[] $^{\rm f}$ For carbon ions independent of particle energy.
\item[] $^{\rm g}$ Not explicitly stated; same data as ICRU49 (1993) -- originally from ICRU37 (1984) -- for protons and alpha particles
\item[] $^{\rm h}$ Not explicitly stated; same composition and elemental data used as ICRU49 (1993) but applied Bragg's additivity rule for mixture
\item[] $^{\rm i}$ For Fig.~B.3 in \cite{IAEA2000}: 79.7\,eV \cite{Hiraroka1995} and 75.3\,eV \cite{Salamon1980}
\item[] $^{\rm j}$ For Fig.~B.3 in \cite{IAEA2000}: 85.9\,eV \cite{Hiraroka1995} and 82.8\,eV\cite{Salamon1980}
\end{indented}
\end{table}
\end{landscape}

\section{Materials and Methods}

\subsection{Stopping power data}
\label{sec:spdata}
\begin{figure}[htb!]
	\centering
	\includegraphics[width=0.7\textwidth]{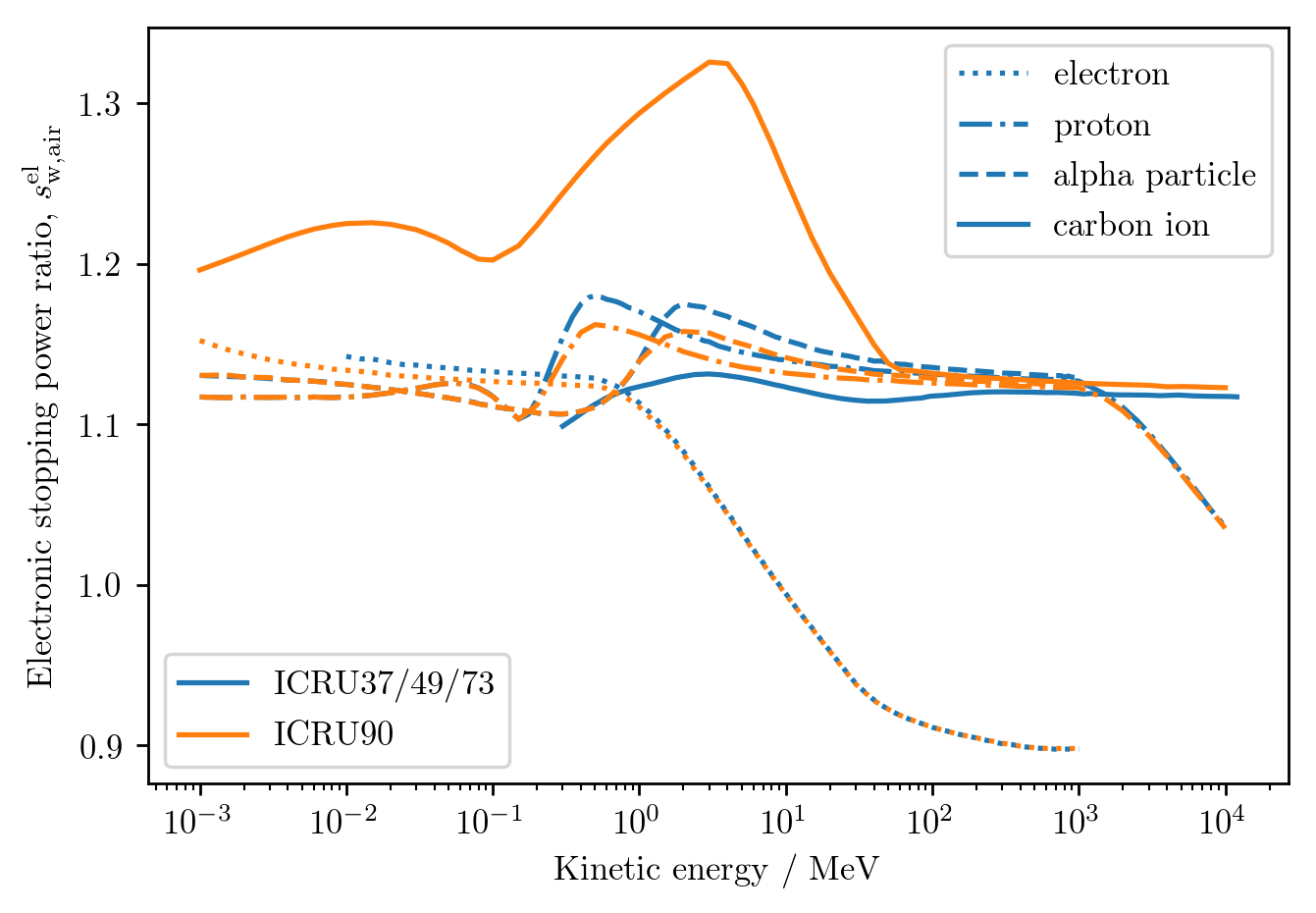}
	\caption{\label{fig:simpleSPR}Water-to-air ratio of monoenergetic electronic stopping power values as a function of kinetic energy. Blue lines correspond to data from former ICRU reports, orange lines to those given in ICRU90.}
\end{figure}

The ICRU90 report contains updated stopping power data for water, graphite and air for electrons, positrons, as well as for protons, alpha particles and carbon ions (Tabs A.1 to A.15 therein). For these three types of ions, electronic, nuclear and total stopping power are given. The electronic stopping-power ratio water to air from ICRU90 data is shown in Fig.~\ref{fig:simpleSPR} in comparison to the data from former reports for electrons, protons and alpha particles\footnote{\url{https://physics.nist.gov/PhysRefData/Star/Text/intro.html}} and the data for carbon ions from the ICRU Report 73 with the Errata (see also Tab.~\ref{tab:quantities}). While changes for electrons, protons and alpha particles are relatively minor, there is a strong increase in stopping-power ratio for carbon ions below 100 MeV. The stopping-power theory and I-values used in ICRU73 differ from those in ICRU49 and ICRU90 which may explain the strong change in the stopping-power ratio for carbon ions. TRS398, on the other hand, predates the publication of ICRU73..\\

ICRU90 does not provide tables for any other lighter ($3\le Z \le 5$) or heavier ($Z>6$) secondary fragments as created by inelastic nuclear scattering of a carbon ion beam in an absorber. Also, the kinetic energy of the ICRU90 tables for alpha particles is limited to 1000\,MeV which does not cover the range of energies for $Z=2$ fragments found in clinical carbon beams with large penetration depths (up to approx. 30\,cm in water corresponding to initial kinetic energy of 430\,MeV/u).

To thus complement the tables and generate the full set of unrestricted electronic stopping power data necessary for this study, we followed closely the approach specified in ICRU90 (Sec.~A.3 therein):
\begin{itemize}
	\item The MSTAR code (v3.12, \cite{PaulAndSchinner2001, PaulAndSchinner2002}) was used to compute data for the low kinetic energy regime below a threshold $T\le T_1$.
	\item In the high energy regime, above a threshold $T\ge T_2$, the BEST code~\cite{BergerPaul1995} was employed\footnote{In the same version as used in ICRU90, i.e. including the update of constants from CODATA 2010. The original code was developed by M.J. Berger and H. Bichsel. The updated BEST code was provided by P. Andreo.}.
	\item To connect the output from both codes in the range $T_1<T<T_2$, $\beta\cdot S_{\rm{el}}(T)/\rho$ was interpolated using a cubic spline.
\end{itemize}
To be consistent with ICRU90, the choice of values for $T_1$ and $T_2$ for the ions not covered in that report were based on the values corresponding to carbon ions as follows:
\begin{itemize}
 \item $T_2$ was set to account for the same ratio $\left<q_1\right>/Z_1 = 0.9522$ of equilibrium charge to nucleus charge obtained for $^{12}$C at 60 MeV.
 \item $T_1$ is set to $0.5 \cdot T_2$.
\end{itemize}
The values for $T_1$ and $T_2$ for lithium to argon ions are provided in Table~\ref{tab:thresholds}. The full set of stopping power tables used in this study, including the additional data, are available in the supplement.

\Table{\label{tab:thresholds}Thresholds for the low (MSTAR code) and high (BEST code) energy regime for stopping power data computation.}
\br
$Z$ & $T_1$ (MeV/nucleon) & $T_2$ (MeV/nucleon) \\
\mr
3 & 0.990 & 1.980 \\
4 & 1.452 & 2.904 \\
5 & 1.960 & 3.919 \\
6 & 2.500 & 5.000 \\
7 & 3.077 & 6.154 \\
8 & 3.682 & 7.363 \\
9 & 4.318 & 8.636 \\
10 & 4.977 & 9.955 \\
11 & 5.665 & 11.330 \\
12 & 6.374 & 12.748 \\
13 & 7.109 & 14.218 \\
14 & 7.864 & 15.728 \\
15 & 8.644 & 17.287 \\
16 & 9.443 & 18.886 \\
17 & 10.265 & 20.531 \\
18 & 11.106 & 22.212 \\
\br
\end{tabular}
\end{indented}
\end{table}

\subsection{Radiation transport code}
\label{sec:Code}
The \textsc{Geant4} toolkit, version 10.3 with patch 1 \cite{Agostinelli2003,Allison2006} was used for radiation transport simulation. It allowed for full implementation of the revised stopping power tables for water and air as given in the ICRU90 report and complementary data generated within this study (see \ref{sec:spdata}). To this end, the \textsc{Geant4} classes \texttt{G4BraggModel}, \texttt{G4BraggIonModel}, \texttt{G4BetheBlochModel} and \texttt{G4IonParametrisedLossModel} which model the energy loss of protons and ions were modified to make explicit use of the new tabulated data for water and air\footnote{The data from ICRU90 report for protons and alpha particles were made available in the \textsc{Geant4} version 10.5, while the data for heavier ions will be included in a future release.}. The modular physics list approach of \textsc{Geant4} was used to account for electromagnetic interactions (physics list \texttt{G4EmStandardPhysics\_option3}), hadronic interactions (physics lists \texttt{G4IonQMDPhysics}, \texttt{G4HadronPhysicsQGSP\_BIC\_HP}, \texttt{G4HadronElasticPhysicsHP} and \texttt{G4StoppingPhysics}) as well as decay physics (physics lists \texttt{G4DecayPhysics} and \texttt{G4RadioactiveDecayPhysics}).

\subsection{Geometry and beam}
\subsubsection{Target}
The target was modeled as a rectangular water volume with lateral extension of $50\times50$\,cm$^2$ and placed in vacuum. In beam direction, the total thickness of the target, i.e. $40$\,cm, was divided into 160 slabs with 0.25\,cm thickness each in order to comply with the use of unrestricted stopping power, see sec.~\ref{sec:IonTransport}. 

\subsubsection{Primary beam}

Two cases were investigated: i) monoenergetic carbon ion beams (no energy spread), and ii) SOBP fields generated by the superposition of 3\,mm-spaced Bragg curves from quasi monoenergetic carbon ion beams (c.f. Sec.~\ref{sec:IDDs} and \ref{sec:SOBPoptim}). In either case, the primary beam was modeled as a thin beam centered on the $z$-axis and traveling in $z+$ direction. The scoring slabs are considered laterally large enough to fully contain the primary beam and the secondary charged particles.

\subsubsection{Depth-dose base data}
\label{sec:IDDs}
A base data set of laterally integrated depth-dose curves representing pristine Bragg peaks with ranges of 3--30\,cm was generated accounting for steps in range of 3\,mm. In contrast to the pure monoenergetic beams, the initial energy spread was here modulated by a 3\,mm ripple filter emulating a clinical carbon-ion beam.

\subsubsection{Computation of biological dose}
\label{sec:RBE}
The depth-dose base data set was complemented with depth curves of $\alpha$ and $\beta$ values for the cell response to the quasi-monoenergetic carbon ion beams following the linear-quadratic model. In order to account for the depth-dependent fluence and energy spectra of carbon ions and secondary charged fragments, the depth curves of $\alpha$ and $\beta$ values were computed in a multi-step process:
\begin{itemize}
  \item First, we computed the cell response to ion irradiation for a series of monoenergetic heavy charged particles $^1$H, $^4$He, $^6$Li, $^8$Be, $^{10}$B, $^{12}$C, $^{14}$N, $^{16}$O in the energy interval from 0.001 to 1000\,MeV/nucleon using the `Compound Poisson Process with Successive Convolution' (CPPSC) model implemented in the \textsc{libamtrack} library~\cite{Greilich2014}. In particular, we assumed as reference condition photon cell response of $\alpha_{\rm X}=0.1$\,Gy$^{-1}$ and $\alpha_{\rm X}/\beta_{\rm X}=2$\,Gy.
  \item Second, the $\alpha^i_{\rm HCP}(T)$ and $\beta^i_{\rm HCP}(T)$ for a heavy charged particle type $i$ at energy $T$ is estimated by linear regression fit of the ion cell response in the dose interval $[0.5,5]$\,Gy using the linear-quadratic model.
  \item Third, depth-dependent $\alpha(z)$ and $\beta(z)$ values for each quasi-monoenergetic carbon ion beam in the base data set were generated by the additivity rules of Zaider and Rossi~\cite{Zaider1980}: 
\begin{equation}\label{Eq:addRuleAlpha}
\alpha(z) = \frac{\sum_{i} \int_0^{\infty}\Phi_{T,i} \cdot \left( S_i(T)/\rho\right)_{\rm{w}} \cdot \alpha^i_{\rm HCP}(T) \cdot dT}{\sum_{i} \int_0^{\infty}\Phi_{T,i} \cdot \left( S_i(T)/\rho\right)_{\rm{w}} \cdot dT}
\end{equation}
and
\begin{equation}\label{Eq:addRuleBeta}
\sqrt{\beta(z)} = \frac{\sum_{i} \int_0^{\infty}\Phi_{T,i} \cdot \left( S_i(T)/\rho\right)_{\rm{w}} \cdot \sqrt{\beta^i_{\rm HCP}(T)}  \cdot dT}{\sum_{i} \int_0^{\infty}\Phi_{T,i} \cdot \left( S_i(T)/\rho\right)_{\rm{w}}  \cdot dT}
\end{equation}

\end{itemize}

\subsubsection{Spread-out Bragg peak optimization}
\label{sec:SOBPoptim}
Spread-out Bragg peaks were composed by weighted superposition of laterally integrated depth-dose curves from the base data set (Sec.~\ref{sec:IDDs}). The weights were determined by minimizing the squared residuals to a set, constant physical (2\,Gy) or biological (3\,GyRBE) dose across the SOBP region using the extension package \textsc{HITXML}, version 0.9.12\footnote{\url{https://r-forge.r-project.org/projects/hitxml/}} for the programming language R\cite{R2018}. For biological dose optimization, the RBE was derived using the $\alpha$ and $\beta$ data for the carbon beam tabulated with depth (Sec.~\ref{sec:RBE}). The resulting $\alpha$ and $\beta$ were obtained by applying the same additivity rules as in Eqs.~\ref{Eq:addRuleAlpha} and \ref{Eq:addRuleBeta}:
\begin{equation}
\alpha(z) = \sum_{k}\frac{d_k(z)}{D(z)}\alpha_k(z)
\end{equation}
and
\begin{equation}
\sqrt{\beta(z)} = \sum_{k}\frac{d_k(z)}{D(z)}\sqrt{\beta_k(z)}
\end{equation}
where $d_k(z)$ is the dose contribution from the $k$-th depth dose curve to the total dose $D(z)$ at a specific depth $z$.

\subsection{Computational procedure for stopping-power ratios}
\subsubsection{General}
Following Bragg-Gray cavity theory, appendix B.6.1 of the TRS398 code of pratice \cite{IAEA2000} defines the fluence-weighted stopping-power ratio as

\begin{equation}
	s_{\rm{w,air}}^{\rm{TRS}}=\frac{\sum_i \int_0^{\infty}\Phi_{T,i} \cdot \left( S_i(T)/\rho\right)_{\rm{w}} \cdot dT}{\sum_i \int_0^{\infty}\Phi_{T,i} \cdot \left( S_i(T)/\rho\right)_{\rm{air}} \cdot dT}
	\label{Eq:sPure}
\end{equation}
where $\Phi_{T,i}$ is the fluence differential in kinetic energy $T$ in water, and $\left(S_i(T)/\rho\right)_{\rm{w}}$ and $\left(S_i(T)/\rho\right)_{\rm{air}}$ are the unrestricted mass stopping powers at energy $T$ in water and air, respectively. The index of summation $i$ includes both primary ions and fragmented nuclei, but no secondary electrons.\\

In Monte Carlo radiation transport, an upper limit of kinetic energy $T_{\rm{max},i}$ which is not exceeded by any particle of type $i$ is set. More importantly, however, a lower limit $T_{\rm{min},i}$ has to be defined below which particle transport is terminated or faded out and the remaining kinetic energy of the track ends is locally deposited. This leads to a modification of Eq.~\ref{Eq:sPure}:

\begin{equation}
s_{\rm{w,air}}^{\rm{MC}}=\frac{\sum_i \int_{T_{\rm{min},i}}^{T_{\rm{max},i}}\Phi_{T,i} \cdot \left( S_i(T)/\rho\right)_{\rm{w}} \cdot dT + D^{\rm{TE}}_{i,\rm{w}}}{\sum_i \int_{T_{\rm{min},i}}^{T_{\rm{max},i}} \Phi_{T,i} \cdot \left( S_i(T)/\rho\right)_{\rm{air}} \cdot dT + D^{\rm{TE}}_{i,\rm{air}}}
\label{Eq:sMC}
\end{equation}
with $D^{\rm{TE}}_i$ being the contribution to dose of the `track-ends' in water and air, respectively. The impact of this lower integration limit and omitting the $D^{\rm{TE}}_i$ terms were discussed in previous studies \cite{Geithner2006, Henkner2009} and transport threshold values $T_{\rm{min}}$ for ions in the order of tens of keV/u were assumed to have negligible impact on the resulting $s_{\rm{w,air}}$.\\

Eq.~\ref{Eq:sPure} considers only ion transport and local deposition of all energy, not considering the dispersion of energy by secondary electrons. This is obtained in the Monte Carlo simulations by setting the cut-off energy for the production of secondary electrons higher than the maximum energy resulting from impact ionization events. This imposes a minimum for the size of geometrical structures such that the mean-chord length in the geometry should not be smaller than the CSDA range of secondary electrons. In this study, the maximum beam energy of 430\,MeV/nucleon sets an upper limit for the energy transferred to secondary electrons to 0.935\,MeV, corresponding to a CSDA range of 0.44\,g/cm$^{2}$. As the mean-chord length for the slabs is approximately twice the slab thickness, the thickness of 0.25\,cm is large enough to not violate the condition above. 

In addition, any directional shift between the point of secondary electron generation and actual dose deposition can be neglected due the very low average energy of the electrons produced by the impact ionization events of carbon ions and secondary nuclear fragments. This is remarkably different to the case of high energy photon beams characterized by a substantially higher energy transferred to secondary electrons.\\

Volume averaging in the slabs occurs especially at high gradients, i.e. close to the Bragg peak. To enable SPR computation with higher spatial resolution, full electron transport and restricted stopping powers have to be employed. However, at least one study showed that this can yield problematic results~\cite{Sanchez2013}. Since conditions for reference dosimetry do not include measurements close to the Bragg peak, this study focused on the use of unrestricted stopping powers and ion-only transport.

\subsubsection{Implementation}
\label{sec:IonTransport}
Stopping-power ratios were computed using Eq.~\ref{Eq:sMC} with $T_{\rm{min}}=1$~keV for all ion types corresponding to the lowest energy in the tabulations available in the ICRU90 report~\cite{ICRU90}. To avoid binning artifacts with respect to $T$, the numerator and denominator in Eq.~\ref{Eq:sMC} were computed `in-flight', i.e. during the simulation. In condensed history (CH) Monte Carlo particle transport, particles are followed in finite CH steps between `catastrophic' events, e.g.~the production of delta electrons or secondary fragments in nucleus-nucleus reactions. Along these CH steps, the particle looses energy continuously in electromagnetic collisions below the production threshold. In this study, the corresponding production threshold was set to 1\,GeV to suppress secondary electron production and allow for the use of unrestricted ion stopping power. For each CH step, therefore, its contribution to the integrands in Eq.~\ref{Eq:sMC}
\begin{equation}
\int^{T_j}_{T_j-\Delta T_j}\left(\Phi_{T',i}\right)_j \cdot \left( S_i(T')/\rho\right) \cdot dT'
\label{Eq:integrandExact}
\end{equation} 
was evaluated by
\begin{equation}
\frac{1}{V} \int_{r_i(T_j-\Delta T_j)}^{r_i(T_j)} S_i(T_i(r'))/\rho \cdot dr'
\label{Eq:integrand}
\end{equation} 
where $r_i(T)$ corresponds to the residual range at kinetic energy $T$ for a particle of type $i$ in the continuous slowing down approximation (CSDA). To implement the computation of Eq.~\ref{Eq:integrand}, first, relations between particle energy $T$ and residual range $r$ were obtained for all particle types $i$ by tabulating the integral of the reciprocal of stopping power data generated according to Sec.~\ref{sec:spdata}
\begin{equation}
r_i(T)=\int_0^T \frac{1}{S_i(T')}\cdot dT'
\label{Eq:rangeEnergy}
\end{equation}
Second, a inverse lookup table was obtained providing $T_i(r)$. These tables were then applied to tabulate
\begin{equation}
\frac{1}{V} \int_{0}^{r_i(T)} S_i(T_i(r'))/\rho \cdot dr'
\end{equation} 
which was eventually used for fast computation of the integrands in Eq.~\ref{Eq:sMC} at each CH step.\\

This numerical procedure corresponds to an extension of the Method 3 for the calculation of fluence differential in energy presented in \cite{Hartmann2017} (cf. Eq.~20 therein). It allows to take the variation of stopping power during the step into account without binning artifacts. This is an advantage over a simple multiplication of the step length $l_j$ by stopping power (corresponding to Method 1 in \cite{Hartmann2017}). In such simplified approach, Eq.~\ref{Eq:integrandExact} would be approximated by $l_j/V\cdot S_i(T_j)/\rho$ where $T_j$ is the energy before, during, or after the step depending of the implementation. This approximation is often used and can be improved by shorter step lengths -- but only at steeply increasing computational costs which will eventually render any refinement infeasible.

The integration
\begin{equation}
\frac{1}{V} \int_{r_i(T_j-\Delta T_j)}^{r_i(T_j)} dr' \approx \int^{T_j}_{T_j-\Delta T_j}\left(\Phi_{T',i}\right)_j \cdot dT' = \left(\Phi_{i}\right)_j
\end{equation} 
yields an approximation of the contribution of step $j$ to the fluence of a particle of type $i$ where the difference between the actual geometrical step length $l_j$ and the CSDA step length $(\Delta r)_j$ is neglected.

When a particle reached the lower limit for transport $T_{{\rm min},i}$, the remaining energy divided by the mass of the current volume was added as track-end contribution, i.e.
\begin{equation}
D^{\rm{TE}}_{i,\rm{w}}=\frac{T_{{\rm min},i}}{\rho_{\rm{w}} \cdot V}
\label{Eq:Cnom}
\end{equation}
to the numerator of Eq.~\ref{Eq:sMC} and 
\begin{equation}
D^{\rm{TE}}_{i,\rm{air}}=\frac{T_{{\rm min},i}}{\rho_{\rm{w}} \cdot V}\cdot\frac{\left(S_i(T_{{\rm min},i})/\rho \right)_{\rm{air}}}{\left(S_i(T_{{\rm min},i})/\rho \right)_{\rm{w}}}
\label{Eq:Cdenom}
\end{equation}
to the denominator. To study the impact of using the electronic instead of the total stopping power, a subset of the simulations were repeated using $S_{{\rm el},i}(T)$ instead of $S_i(T)$ in Eqs.~\ref{Eq:sMC}, \ref{Eq:Cnom}, and~\ref{Eq:Cdenom}.

\subsection{Beam quality specifier}
Instead of the initial kinetic energy, range or SOBP width as a beam quality specifier, we use the residual range $R_{\rm{res}}$ at a depth $z$
\begin{equation}
R_{\rm{res}} = R_p - z
\end{equation}
where $R_{\rm{p}}$ is the practical range of the beam, i.e. the depth at which the absorbed dose beyond the Bragg peak or spread-out Bragg peak decreases to 10\,\% of its maximum value. Due to the fragmentation tail, the Bragg peak may not directly drop to a 10\,\% level. In this case a tangent at the steepest point of the distal fall-off is used to construct the virtual position of $R_{\rm{p}}$.

\subsection{Beam quality correction factors}
\label{sec:PerturbationFactors}
$k_Q$ factors were computed according to Eq.~\ref{eq:kq}, using for the ion-related quantities in the numerator:
\begin{itemize}
	\item The stopping-power ratio values for carbon beams obtained in this study, 
	\item the updated, constant $W/e$ value from ICRU90 (34.71\,eV, Tab.~\ref{tab:quantities}) and 
	\item the recommended total, chamber-independent perturbation factor $p_Q=1$ (\emph{c.f}. Tab.~\ref{tab:quantities}).
\end{itemize}

For the $^{60}$Co related terms in Eq.~\ref{eq:kq}, a $W/e$ value of 33.97\,eV was used. The procedure for the two remaining quantities depended on the availability of data for perturbation factors:
\begin{enumerate}
	\item Gom\`a \emph{et al.}~\cite{Goma2016} provide combined data for $\left(s_{\rm{w,air}}\cdot p\right)_{^{60}{\rm{Co}}}$ which were directly inserted for the ionization chamber types studied therein.
	\item Muir \emph{et al.}~\cite{Muir2010} presented updated perturbation factors for a number of chambers which are used in the draft of the German Code of Practice (DIN6801-1)\cite{DIN2016}. These were directly used for computation together with the $\left(s_{\rm{w,air}}\right)_{^{60}{\rm{Co}}}$ value of 1.127 from ICRU90. 
	\item Perturbation factors extracted from TRS398 and perturbation factors in DIN6801-1 reported to be identical to TRS398 were multiplied by 1.012 to follow the ICRU90 recommendation (Tab.~\ref{tab:quantities}) together with $\left(s_{\rm{w,air}}\right)_{^{60}{\rm{Co}}}=1.127$.
\end{enumerate}

\begin{landscape}
\Table{\label{tab:situations}Irradiation situations modelled.}
\br
Section & Class & Range & Simulation & Thresholds\\
\br
& & & Ion transport only, &  \\
\ref{sec:Precursor} & Pristine Bragg peaks & 12.8\,cm, 3--30\,cm & unrestricted total / electronic  & $T_{\rm{min}} = 1...500$\,keV\\
& & & stopping power & \\
\mr
& Pristine Bragg peaks & 3--30\,cm & & \\
& & & & \\
& & 8\,cm, width 2--4\,cm & Ion transport only, & \\
\ref{sec:Ions} & Physically optimized SOBPs & 16\,cm, width 2--12\,cm & unrestricted total & $T_{\rm{min}} = 1$\,keV\\
& & 30\,cm, width 2--12\,cm & stopping power & \\
& & & & \\
& Biologically optimized SOBPs & 16\,cm, width 2--12\,cm & & \\
\br
\end{tabular}
\end{indented}
\end{table}
\end{landscape}

\section{Results and Discussion}
Tab.~\ref{tab:situations} gives an overview on the beam configurations and simulation parameters used in this study. After confirming the soundness of the approach used for the computation of SPR (Sec.~\ref{sec:Precursor}), the results for a wide range of reference conditions are given in Sec.~\ref{sec:Ions}. Eventually, these are used to obtain updated $k_Q$ factors for a number of ionization chambers (Sec.~\ref{sec:kQ}).

\subsection{General}
\label{sec:Precursor}

\subsubsection{Comparison to previous results}
\begin{figure}[htb!]
	\centering
	\includegraphics[width=0.7\textwidth]{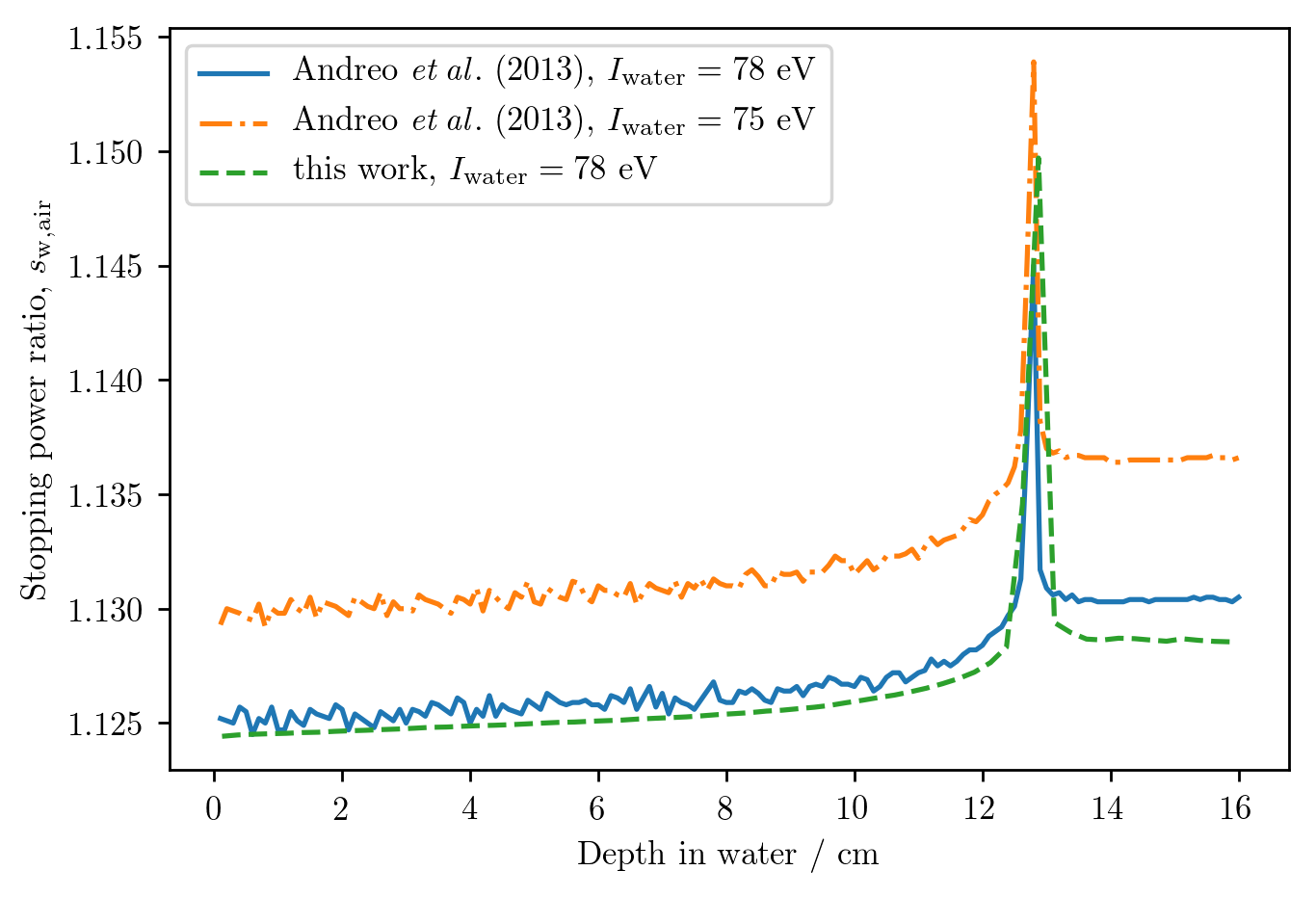}
	\caption{\label{fig:andreo}Stopping-power ratio water to air, $s_{\rm{w,air}}$, for a monoenergetic 250\,MeV/u carbon ion beam. Solid and dash-dotted lines correspond to data from Andreo \emph{et al.}~\cite{Andreo2013}, dashed line to those obtained in this study.}
\end{figure}

Andreo \emph{et al.}~\cite{Andreo2013} estimated the impact of the new key data recommendation, esp. the change in $I$ for water, for a monoenergetic carbon beam of initial kinetic energy of 250\,MeV/u using the \textsc{Shield-HIT} transport simulation code v10. The SPR values are 0.5\,\% lower for $I_{\rm{water}}=78$\,eV in comparison to $I_{\rm{water}}=75$\,eV as used in the TRS398 code of practice (Fig.~\ref{fig:andreo}). These results were compared to the outcome from the modified \textsc{Geant4} code used in this study (cf. Sec. \ref{sec:Code}) which agrees within 0.1\,\% with the data from~\cite{Andreo2013} for $I_{\rm{water}}=78$\,eV, except in the immediate vicinity of and beyond the Bragg peak. In particular, the observed differences in the tail are largely related to the different yield of secondary nuclear fragments generated by each Monte Carlo code.

\subsubsection{Impact of lower integration threshold}
\begin{figure}[htb!]
	\centering
	\includegraphics[width=0.7\textwidth]{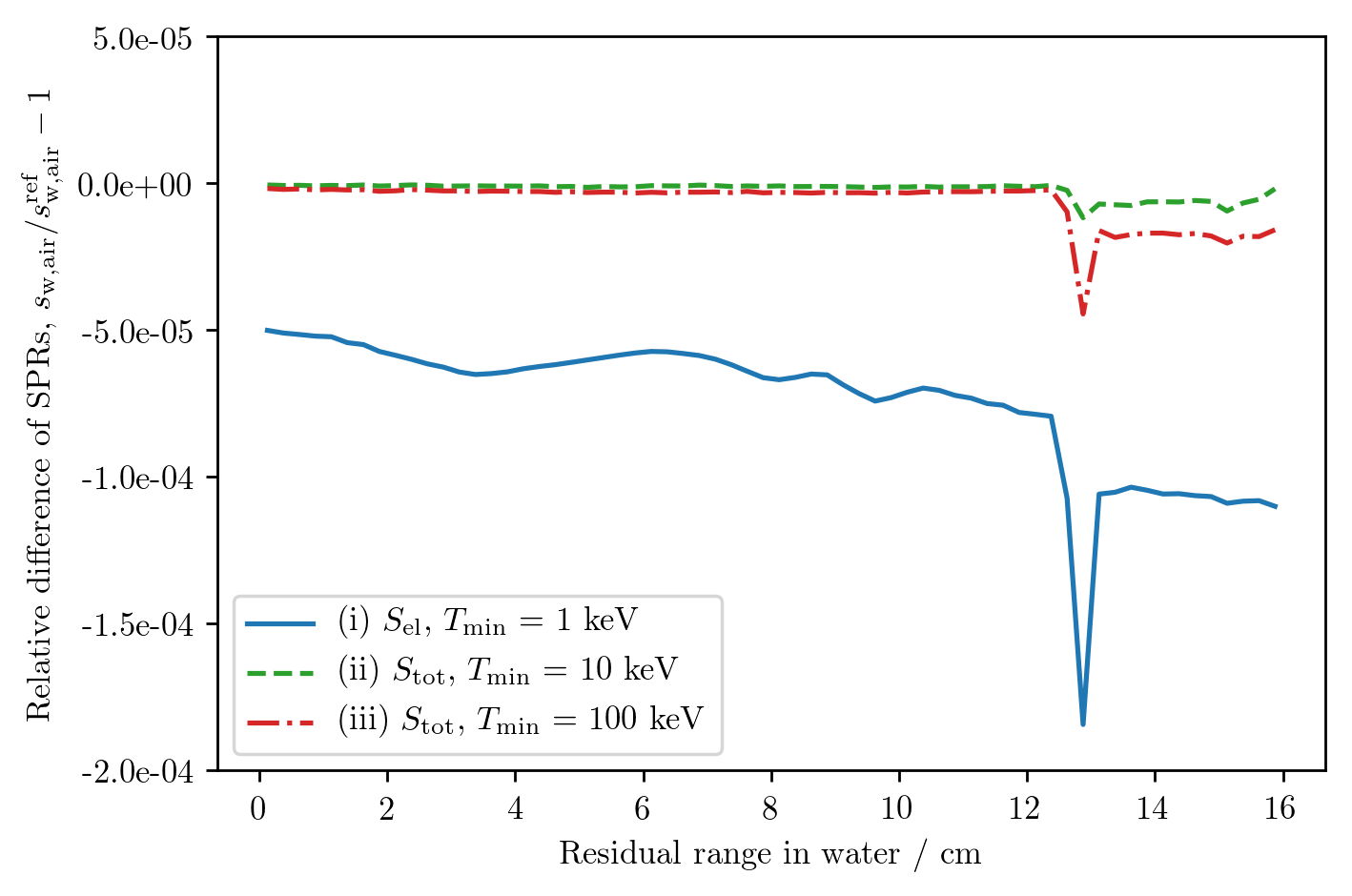}
	\caption{\label{fig:Tmin}Relative difference of $s_{\rm{w,air}}$ to data from this study ($s_{\rm{w,air}}^{\rm{ref}}$) shown in Fig.~\ref{fig:andreo} using (i) the electronic stopping power and $T_{\rm{min}}=1$\,keV, and (ii) total stopping power and lower integration threshold $T_{\rm{min}}$=10\,keV, and (iii) total stopping power and lower integration threshold $T_{\rm{min}}$=100\,keV.}
\end{figure}

Fig.~\ref{fig:Tmin} shows that the impact of a variation of the lower integration threshold, $T_{\rm{min}}$,  between 1 and 100\,keV on the calculation of stopping-power ratio for the carbon ion beam found in Fig.~\ref{fig:andreo} is negligible. Similar results were observed for pristine Bragg peaks with residual ranges between 3 and 30\,cm and lower integration thresholds between 1 and 500\,keV (data not shown).

\subsubsection{Total vs.~electronic stopping power}
The impact of using electronic stopping power versus total stopping power to calculate the stopping-power ratio is also shown in Fig.~\ref{fig:Tmin}. A systematically smaller stopping-power ratio was observed when using the electronic stopping power but the effect is in the order of ($0.5-2\cdot10^{-4}$) and does thus not play a role in this study. Similar results were obtained for residual range in water between 3 and 30\,cm (data not shown).

\subsection{Stopping-power ratio computations}
\label{sec:Ions}

\subsubsection{Pristine Bragg peaks}
\begin{figure}[htb!]
	\centering
	\includegraphics[width=0.7\textwidth]{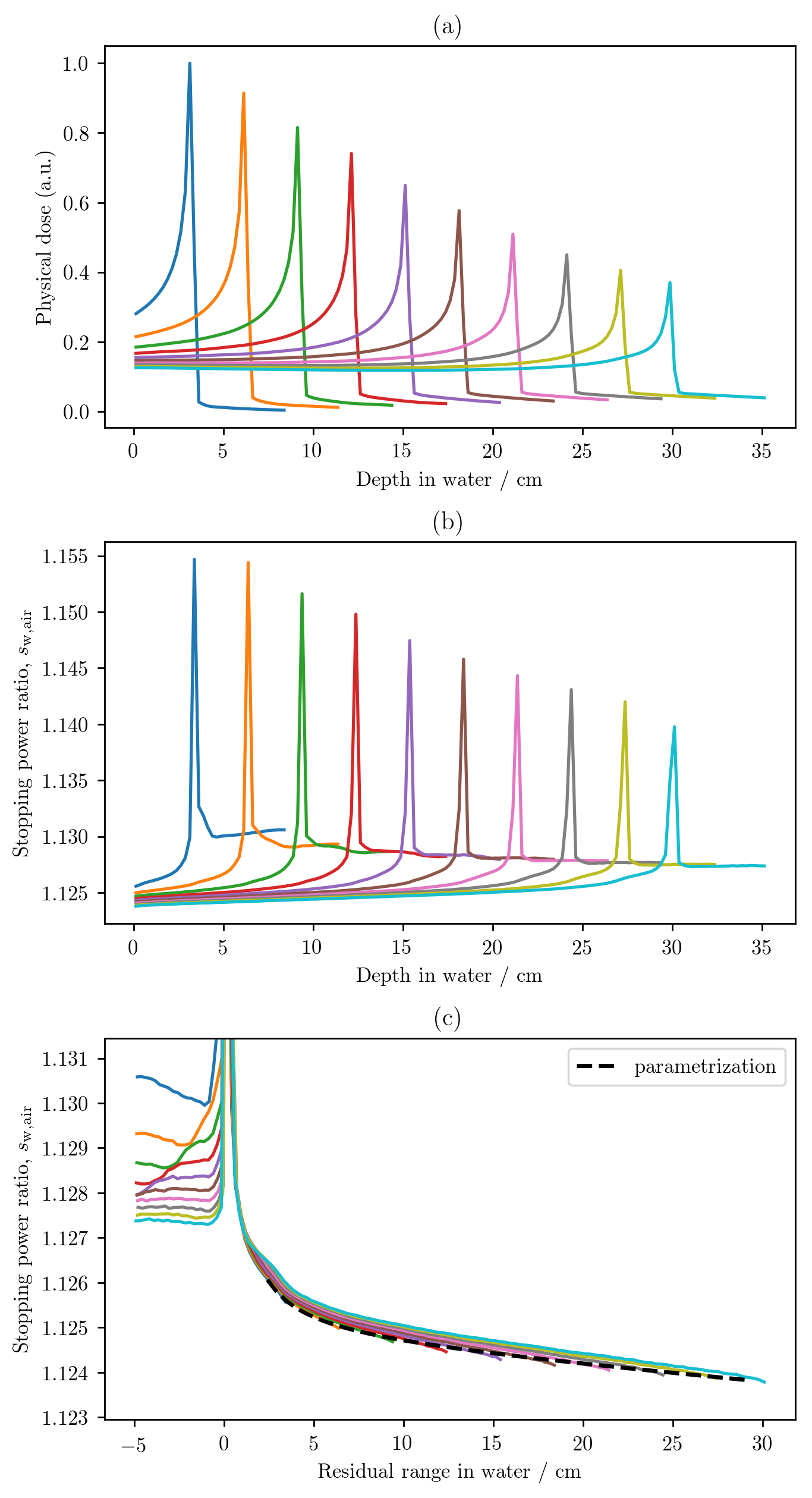}
	\caption{\label{fig:mono}(a) Depth-dose profile for monoenergetic carbon ion beams with range in water between 3 and 30\,cm, and stopping-power ratio, $s_{\rm{w,air}}$, as a function of (b) depth and (c) residual range in water. The curve fit to the stopping-power ratio as a function of range in water using Eq.~\ref{Eq:fitSPR} is shown by the dashed line.}
\end{figure}

Depth-dose profiles and stopping-power ratios obtained for monoenergetic carbon ion beams with residual range in water from 3 to 30\,cm are shown in Fig.~\ref{fig:mono}. High values of $s_{\rm{w,air}}$ are observed around the Bragg peak for beams with small residual ranges. However, as the effect of energy straggling increases for beams with larger range -- as seen in the increasing width of the Bragg peak -- the spatial concentration of stopping carbon ions for which $s_{\rm{w,air}}$ is large (see Fig.~\ref{fig:simpleSPR}) decreases and the high stopping-power ratio values fade away.\\

\begin{figure}[htb!]
	\centering
	\includegraphics[width=0.7\textwidth]{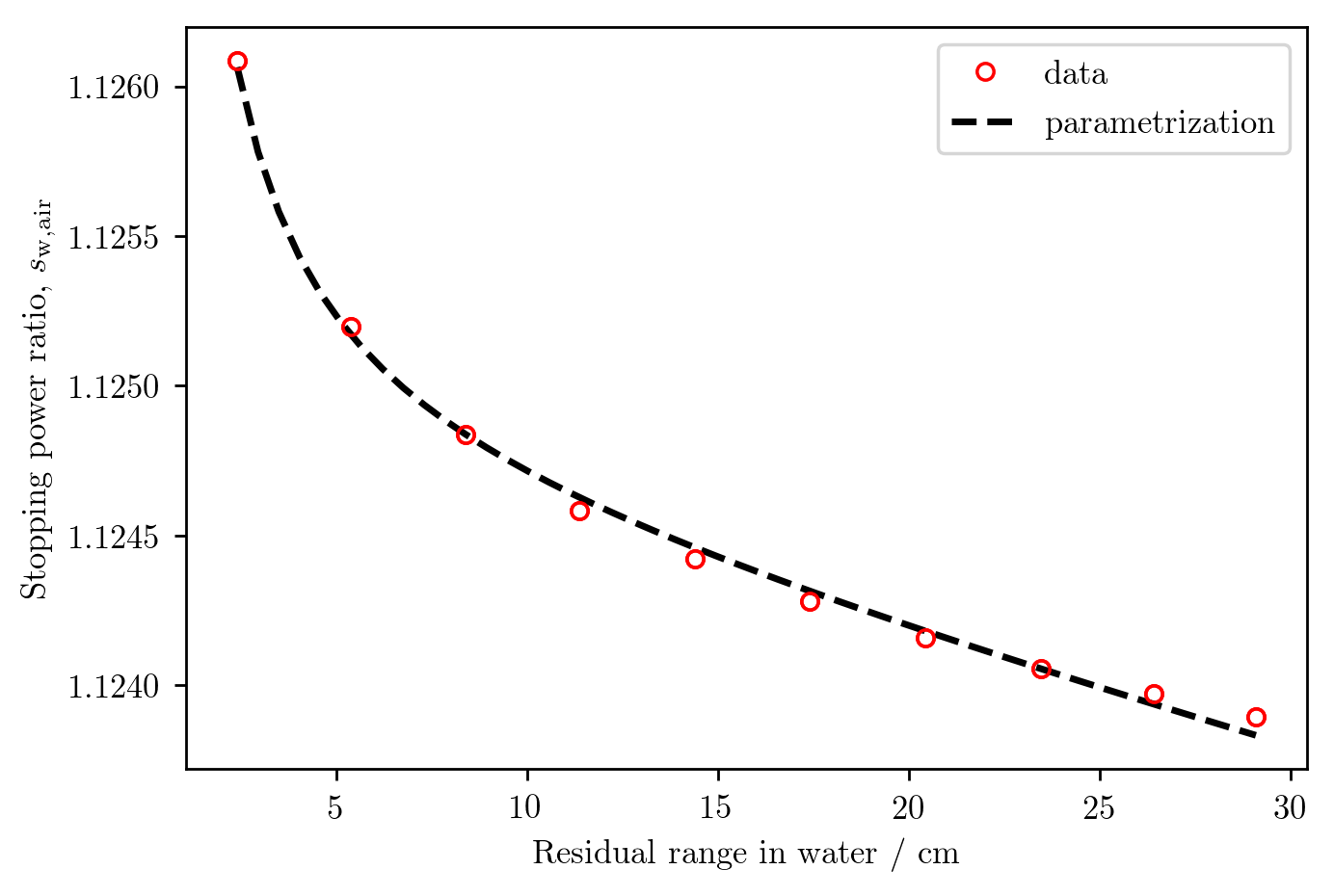}
	\caption{\label{fig:monoFit}Results of fitting the function of Eq.~\ref{Eq:fitSPR} to the data points for $s_{\rm{w,air}}$ in pristine Bragg peaks in a depth of 1\,g\,cm$^{-2}$ in water. The parameters of the fit are presented in Tab.~\ref{tab:param}.}
\end{figure}

In Fig.~\ref{fig:monoFit}, the $s_{\rm{w,air}}$ values needed for reference dosimetry at a depth of 1\,g\,cm$^{-2}$ for the beam qualities investigated were parametrized as a function of residual range $R_{\rm{res}}$:
\begin{equation}\label{Eq:fitSPR}
	s_{\rm{w,air}}(R_{\rm{res}}) = a + b\cdot R_{\rm{res}} + \frac{c}{R_{\rm{res}}}\quad.
\end{equation}
Tab.~\ref{tab:param} gives the resulting values for the parameters. To simplify the interpretation and usage, the curve was fitted to the data points with the constraint
\begin{equation}\label{Eq:fitSPRconstraint}
	c = -b\cdot \left(R_{\rm{res}}^{\rm{rep}}\right)^2
\end{equation}
with $R_{\rm{res}}^{\rm{rep}}=10$\,cm. At this depth, $a$ equals $1.1247$ and represents $s_{\rm{w,air}}$ within an interval (-0.07\,\%,+0.12\,\%) for beams with residual ranges in water between 3 and 30\,cm. 

\subsubsection{Spread-out Bragg peaks}
\begin{figure}[htb!]
	\centering
	\includegraphics[width=0.9\textwidth]{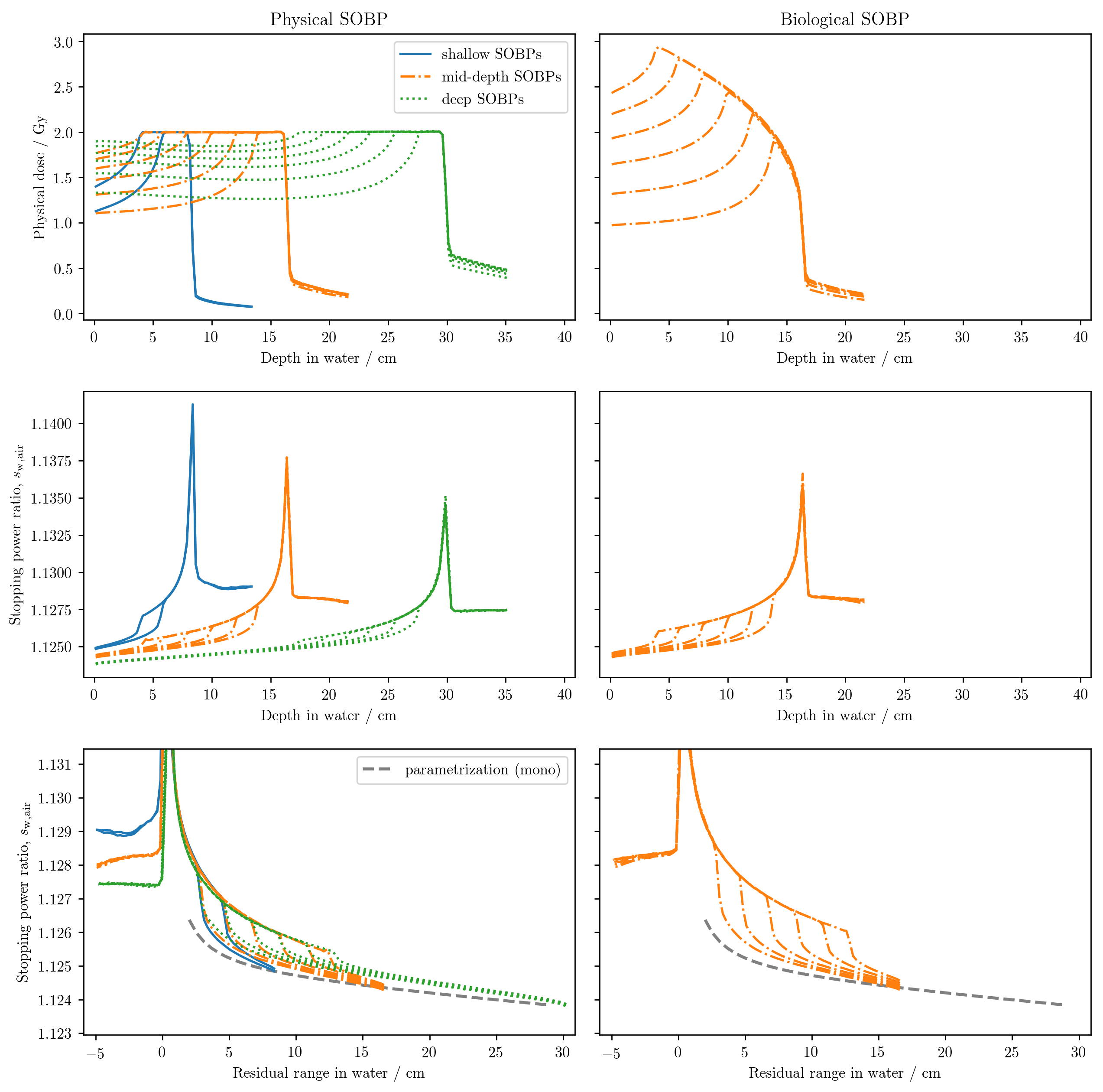}
	\caption{\label{fig:phySOBP}(Left side) Physical depth-dose profiles for SOBPs of short, mid, and large range and of different widths (upper panel), and stopping-power ratio, $s_{\rm{w,air}}$, as a function of depth (middle) and  residual range in water (lower panel). (Right side) Same data, but for biologically optimized SOBPs, mid range only. For better comparison, the function fitted to the data points for pristine Bragg curves (Figs.~\ref{fig:mono} and \ref{fig:monoFit}) is given in the lower panels.}
\end{figure}

The depth-dose profiles and stopping-power ratios obtained from both physically and biologically optimized SOBPs with varying width and range are shown in Fig.~\ref{fig:phySOBP}. A pattern can be observed in the stopping-power ratio characterized by an abrupt increase of $s_{\rm{w,air}}$ at the proximal edge of the SOBP of about 2\,\textperthousand{} due to the higher stopping-power ratio of stopping carbon ions present in this part. The $s_{\rm{w,air}}$ values within the high dose region for different SOBP widths lie well on top of one another (middle panels). This is also observed for SOBPs at different depths when analyzed as a function of residual range (lower panels). This supports the suitability of $R_{\rm{res}}$ as a simplified beam quality specifier.\\

\begin{figure}[htb!]
	\centering
	\includegraphics[width=0.9\textwidth]{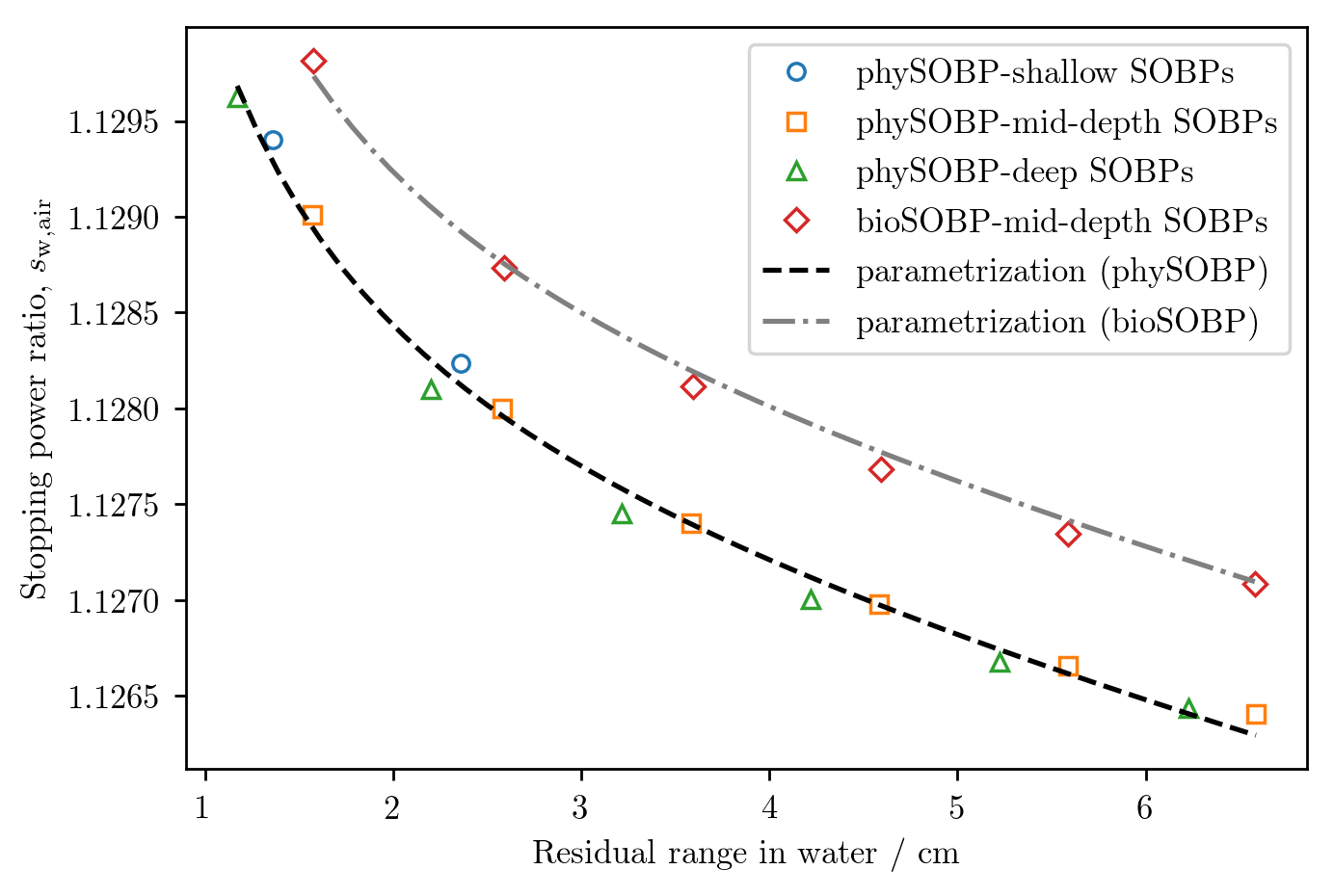}
	\caption{\label{fig:SOBPfit}Stopping power ratio, $s_{\rm{w,air}}$ at the mid-SOBP position, as a function of residual range in water for physical and biological SOBPs. The parametrizations of the stopping-power ratio as a function of residual range in water using Eq.~\ref{Eq:fitSPR} are shown by the dashed and dashed-dotted lines for the physical and the biological SOBP, respectively.}
\end{figure}

In the same way as done for pristine Bragg curves, $s_{\rm{w,air}}(R_{\rm{res}})$ was parameterized by Eq.~\ref{Eq:fitSPR} for reference conditions, i.e. the center of the SOBPs. The parameterization and the fit parameters are shown in Fig.~\ref{fig:SOBPfit} and Tab.~\ref{tab:param}, respectively. The shape of the function is different than for pristine Bragg peaks as reflected in the values of $b$ and $c$. The data points at the mid-SOBP for the physical SOBPs were fitted using the constraint in Eq.~\ref{Eq:fitSPRconstraint} with $R_{\rm{res}}^{\rm{rep}}=3.5$\,cm. Here $a=1.1274$ represent $s_{\rm{w,air}}$ within (-0.09\,\%,+0.18\,\%) for the different widths and depths of the physical SOBPs investigated in this study.\\

For the case of the mid-SOBP positions for the biological SOBPs, the same parameters $b$ and $c$ were used, and only the parameter $a$ was fitted, accounting for a systematic shift of the $s_{\rm{w,air}}$ values. Since the weights of the most distal quasi-monoenergetic beams contributing to the SOBP is reduced by taking into account a higher RBE in biological optimization, the stopping power values in the SOBP region systematically increases. Reference dosimetry should focus on the determination of physical dose and corresponding beam configurations. However, $s_{\rm{w,air}}$ increases only by about 0.8\,\textperthousand~even for a strong deviation of dose from a physically optimized SOBP by an approximate factor of 2 at the distal end as in the case shown.\\ 

\Table{\label{tab:param}Paramerization for the stopping-power ratio as a function of residual range for reference conditions.}
\br
Class &  Position & $R_{\rm{res}}^{\rm{rep}}$ & $a$ & $b$ & $c$\\
\mr
Pristine Bragg peak &  1\,g\,cm$^{-2}$ depth & 10\,cm & $1.1247$ & $-3.444\cdot10^{-5}$& $3.444\cdot10^{-3}$\\
Physical SOBPs &  Mid-SOBP & 3.5\,cm & $1.1274$ & $-2.418\cdot10^{-4}$& $2.962\cdot10^{-3}$ \\
Biological SOBPs &  Mid-SOBP & 3.5\,cm & $1.1282$ & $-2.418\cdot10^{-4}$& $2.962\cdot10^{-3}$ \\
\br
\end{tabular}
\end{indented}
\end{table}

\subsubsection{Choice of SPR values}
\label{sec:choiceSPR}
The value of $s_{\rm{w,air}}$ suggested by these studies for use in reference dosimetry can be summarized as follows:
\begin{itemize}
	\item If a single constant $s_{\rm{w,air}}$ is used as recommended in TRS398, an average between the values representative for pristine and for (physically optimized) SOBP configurations should be used, which is 1.126. All values obtained in this study fall into a ($-0.2\,\%$,$+0.3\,\%$) interval around this value. The change of $-0.4\,\%$ with respect to the recommended value of 1.130 in TRS398 corresponds with the statements given in ICRU90. 
	\item If the SPR should be representative for either a pristine or SOBP situation, then the corresponding value of $a$ from Tab.~\ref{tab:param} can be used.   
	\item Eventually, the full parameterization given in Tab.~\ref{tab:param} can be employed to study the dependence of $s_{\rm{w,air}}$ on the specific beam situation.
\end{itemize}

\subsection{Beam quality correction factors}
\label{sec:kQ}

Tab.~\ref{tab:kQ} lists the perturbation factors used for computation and the resulting $k_Q$ data plus experimental data from~\cite{Osinga2017,Osinga2018}. Fig.~\ref{fig:kQDiff} shows a subset excluding chambers for which perturbation factors are taken only from TRS398 (which resulted in a constant factor) and no experimental $k_Q$ data were available. 

\subsubsection{Original values}
Original $k_Q$ data from DIN6801-1 (orange squares) are lower for most chambers than the values from TRS398. A baseline difference of $-0.9\,\%$ arises from the lower SPR for ions (1.121 for a residual range of 15\,cm) used in the German Code of Practice. Experimental values for cylindrical chambers are closer to the TRS398 data (average deviation of $-0.5\,\%$) than to those from DIN6801-1 ($+0.9\,\%$). For plate-parallel chambers, there is only one value for TRS398 ($-0.4\,\%$) but three for DIN6801-1 ($-0.1\,\%$).

\subsubsection{Updated values}
\begin{figure}[htb!]
	\centering
	\includegraphics[width=0.99\textwidth]{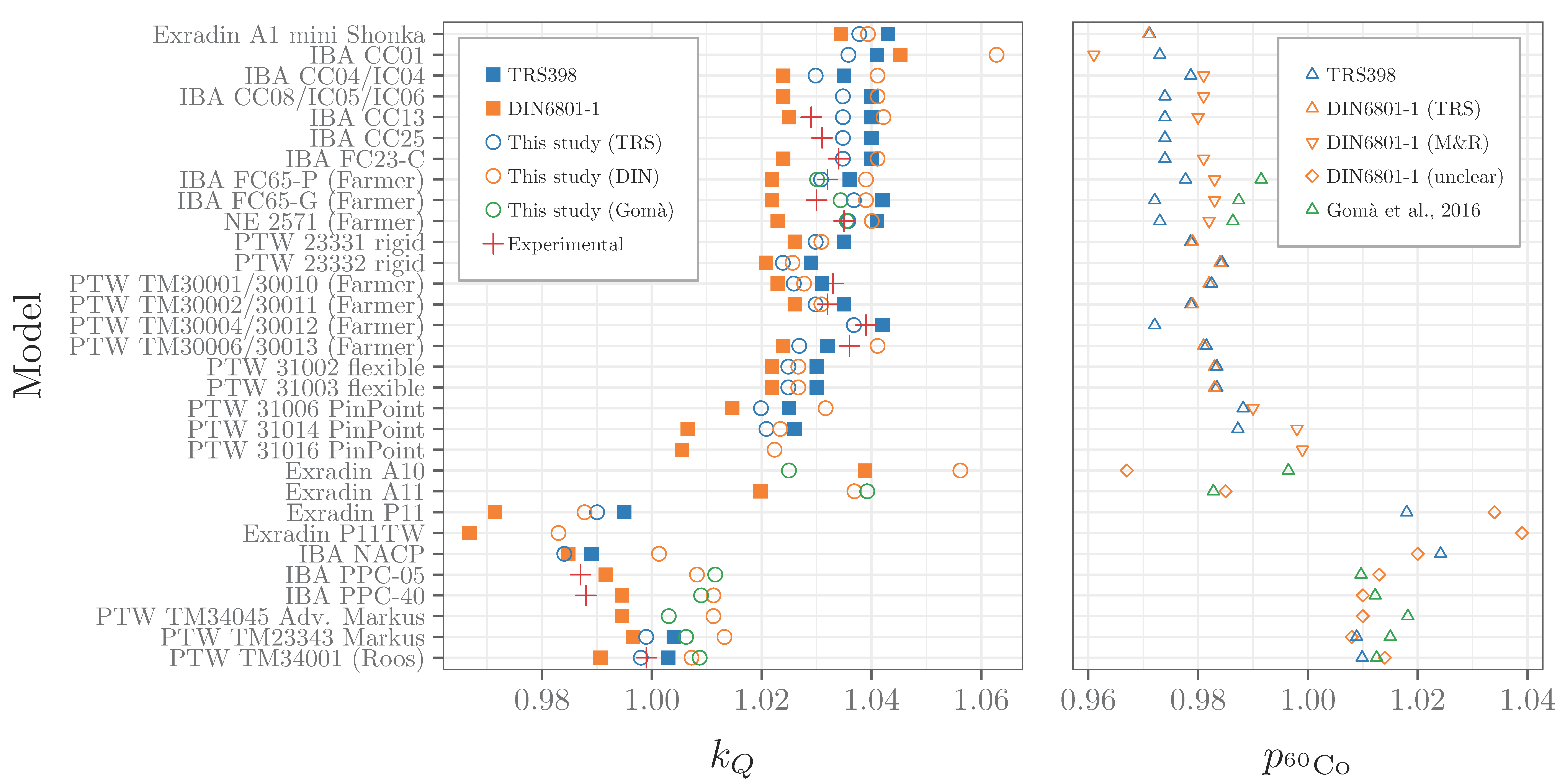}
	\caption{\label{fig:kQDiff}Original and updated beam quality correction factors from TRS398, DIN6801-1 together with updated data from this study and experimental data (left). In the right panel, the corresponding perturbation factors extracted from different sources for the update of $k_Q$ values are shown.}
\end{figure}

For updated $k_Q$ data with perturbation factors taken from TRS398 directly (case i in sec~\ref{sec:PerturbationFactors}), a constant change of $+0.5\,\%$ is seen for all chambers between TRS398 and the recalculations in this study due to the following (cf. Table 7.2 in \cite{ICRU90}):
\begin{itemize} 
	\item The impact of the new key data on the stopping-power ratio water to air for $^{60}$Co is $-0.5\,\%$.
	\item The general chamber perturbation factor sees a large increase from recent Monte Carlo particle transport calculations ($+1.2\,\%$) which yields an overall change of $\left(s_{\rm{w,air}}\cdot p\right)_{{^{60}\rm{Co}}}$ of $+0.7\,\%$ in the denominator of Eq.~\ref{eq:kq}.
	\item While the recommended value for $s_{\rm{w,air}}$ for carbon changes from 1.130 to 1.126 ($-0.4\,\%$), $W/e$ for air changes by $+0.6\,\%$ -- yielding together $+0.2\,\%$ for the numerator. Together with the changes in the $^{60}$Co-related quantities, this totals to $-0.5\,\%$ and is reflected by the nearly constant distance between the corresponding symbols (blue squares, blue circles) in Fig.~\ref{fig:kQDiff}.
\end{itemize}
Interestingly, this compensates the difference to experimental data for cylindrical chambers.\\

For case (ii), i.e. the usage of perturbation factors from DIN6801-1, those updated $k_Q$ values for which perturbation factors were reported to be the same as in TRS398 are very close ($+0.2\,\%$) but not identical to the updated $k_Q$ derived from TRS398 directly. This could be due to round-off errors. Most perturbation factors from Muir and Rogers as used by DIN6801-1, however, are considerably larger than those originally used in TRS398. While this is in principle in line with ICRU90 recommendation of $+1.2\,\%$, the actual magnitude is smaller for most chambers.

The relatively few updated $k_Q$ values based on the combined $\left(s_{\rm{w,air}}\cdot p\right)_{{^{60}\rm{Co}}}$ data from Gom\`a \emph{et al.}, i.e. case (iii), were very close to the data updated from TRS398 directly for cylindrical chambers. For plate-parallel chambers, discrepancies in the order of 1\,\% arise. With the low number of data points currently available, however, no clear conclusion can be drawn. It is noticeable, however, that the experimental data for two chambers (IBA PPPC-05 and -40) just agree with the original $k_Q$ values from DIN6801-1 within the lower bound of their uncertainty ($1.1\,\%$), but other data were generally considerably underestimated by the DIN CoP.\\

\begin{landscape}
	
	\Table{\label{tab:kQ}Pertubation and beam quality correction factors for carbon ion beams. Data for $p_{^{60}\rm{Co}}$ are obtained from \cite{IAEA2000,DIN2016,Goma2016}. Beside $k_Q$ factors from these publications and experimental data from \cite{Osinga2017,Osinga2018}, the updated values from this study using the pertubation factors from the three sources are listed. Factors from DIN6801-1 are expressed in the TRS398 formalism, i.e. including the displacement correction factor $k_r$ in $k_Q$.}
	\br

    & & $p_{^{60}\rm{Co}}$ & & & & & $k_Q$ & & \\
   Chamber type	& \scriptsize{TRS398} & \scriptsize{DIN6801-1} & \scriptsize{Gom\`a \emph{et al.}} & \scriptsize{TRS398} & \scriptsize{DIN6801-1} & & \scriptsize{This study} & & \scriptsize{Experiment} \\
& \tiny{(2006)} & \tiny{(2010)} & \tiny{(2016)} & \tiny{(2006)} & \tiny{(2010)} & \tiny{$p$ TRS398} & \tiny{$p$ DIN} & \tiny{$p$ Gom\`a} & \\
	\mr
	\textit{Cylindrical chambers} & & & & & & & & & \\

Capintec PR-05 mini & 0.969 &  --- &  --- & 1.045 &  --- & 1.040 &  --- &  --- &  --- \\
Capintec PR-06C/G Farmer & 0.977 &  --- &  --- & 1.037 &  --- & 1.032 &  --- &  --- &  --- \\
Exradin A2 Spokas  & 0.960 &  --- &  --- & 1.055 &  --- & 1.050 &  --- &  --- &  --- \\
Exradin T2 Spokas  & 0.995 &  --- &  --- & 1.018 &  --- & 1.013 &  --- &  --- &  --- \\
Exradin A1 mini Shonka  & 0.971 & 0.971 &  --- & 1.043 & 1.034 & 1.038 & 1.052 &  --- &  --- \\
Exradin T1 mini Shonka  & 1.006 &  --- &  --- & 1.007 &  --- & 1.002 &  --- &  --- &  --- \\
Exradin A12 Farmer & 0.972 &  --- &  --- & 1.042 &  --- & 1.037 &  --- &  --- &  --- \\
Far West Tech  IC-18 & 1.007 &  --- &  --- & 1.006 &  --- & 1.001 &  --- &  --- &  --- \\
FZH TK 01 & 0.982 &  --- &  --- & 1.031 &  --- & 1.026 &  --- &  --- &  --- \\
IBA CC01 & 0.973 & 0.961 &  --- & 1.041 & 1.045 & 1.036 & 1.063 &  --- &  --- \\
IBA CC04/IC04 & 0.979 & 0.981 &  --- & 1.035 & 1.024 & 1.030 & 1.041 &  --- &  --- \\
IBA CC08/IC05/IC06 & 0.974 & 0.981 &  --- & 1.040 & 1.024 & 1.035 & 1.041 &  --- &  --- \\
IBA CC13 & 0.974 & 0.980 &  --- & 1.040 & 1.025 & 1.035 & 1.042 &  --- & 1.029 \\
IBA CC25 & 0.974 &  --- &  --- & 1.040 &  --- & 1.035 &  --- &  --- & 1.031 \\
IBA FC23-C & 0.974 & 0.981 &  --- & 1.040 & 1.024 & 1.035 & 1.041 &  --- & 1.034 \\
IBA FC65-P (Farmer) & 0.978 & 0.983 & 0.991 & 1.036 & 1.022 & 1.031 & 1.039 & 1.030 & 1.032 \\
IBA FC65-G (Farmer) & 0.972 & 0.983 & 0.987 & 1.042 & 1.022 & 1.037 & 1.039 & 1.034 & 1.030 \\
Nuclear Assoc 30-750 & 0.979 &  --- &  --- & 1.035 &  --- & 1.030 &  --- &  --- &  --- \\
Nuclear Assoc 30-749 & 0.975 &  --- &  --- & 1.039 &  --- & 1.034 &  --- &  --- &  --- \\
Nuclear Assoc 30-744 & 0.975 &  --- &  --- & 1.039 &  --- & 1.034 &  --- &  --- &  --- \\
Nuclear Assoc 30-716 & 0.975 &  --- &  --- & 1.039 &  --- & 1.034 &  --- &  --- &  --- \\

	\br
\end{tabular}
\end{indented}
\end{table}

\addtocounter{table}{-1}

\Table{Pertubation and beam quality correction factors for carbon ion beams (\textit{continued}).}
\br

& & $p_{^{60}\rm{Co}}$ & & & & & $k_Q$ & & \\
Chamber type	& \scriptsize{TRS398} & \scriptsize{DIN6801-1} & \scriptsize{Gom\`a \emph{et al.}} & \scriptsize{TRS398} & \scriptsize{DIN6801-1} & & \scriptsize{This study} & & \scriptsize{Experiment} \\
& \tiny{(2006)} & \tiny{(2010)} & \tiny{(2016)} & \tiny{(2006)} & \tiny{(2010)} & \tiny{$p$ TRS398} & \tiny{$p$ DIN} & \tiny{$p$ Gom\`a} & \\
\mr

Nuclear Assoc 30-753 Farmer shortened & 0.974 &  --- &  --- & 1.040 &  --- & 1.035 &  --- &  --- &  --- \\
Nuclear Assoc 30-751 Farmer & 0.978 &  --- &  --- & 1.036 &  --- & 1.031 &  --- &  --- &  --- \\
Nuclear Assoc 30-752 Farmer & 0.972 &  --- &  --- & 1.042 &  --- & 1.037 &  --- &  --- &  --- \\
NE 2515 & 0.982 &  --- &  --- & 1.032 &  --- & 1.027 &  --- &  --- &  --- \\
NE 2515/3 & 0.973 &  --- &  --- & 1.041 &  --- & 1.036 &  --- &  --- &  --- \\
NE 2577 & 0.973 &  --- &  --- & 1.041 &  --- & 1.036 &  --- &  --- &  --- \\
NE 2505 Farmer & 0.982 &  --- &  --- & 1.032 &  --- & 1.027 &  --- &  --- &  --- \\
NE 2505/A Farmer & 0.994 &  --- &  --- & 1.019 &  --- & 1.014 &  --- &  --- &  --- \\
NE 2505/3, 3A Farmer & 0.973 &  --- &  --- & 1.041 &  --- & 1.036 &  --- &  --- &  --- \\
NE 2505/3, 3B Farmer & 0.990 &  --- &  --- & 1.023 &  --- & 1.018 &  --- &  --- &  --- \\
NE 2571 (Farmer) & 0.973 & 0.982 & 0.986 & 1.041 & 1.023 & 1.036 & 1.040 & 1.035 & 1.035 \\
NE 2581 Farmer  & 0.995 &  --- &  --- & 1.018 &  --- & 1.013 &  --- &  --- &  --- \\
NE 2561 / 2611 Sec Std & 0.976 &  --- &  --- & 1.038 &  --- & 1.033 &  --- &  --- &  --- \\
PTW 23323 micro & 0.987 &  --- &  --- & 1.026 &  --- & 1.021 &  --- &  --- &  --- \\
PTW 23331 rigid & 0.979 & 0.979 &  --- & 1.035 & 1.026 & 1.030 & 1.043 &  --- &  --- \\
PTW 23332 rigid & 0.984 & 0.984 &  --- & 1.029 & 1.021 & 1.024 & 1.038 &  --- &  --- \\
PTW 23333  & 0.982 &  --- &  --- & 1.031 &  --- & 1.026 &  --- &  --- &  --- \\
PTW TM30001/30010 (Farmer) & 0.982 & 0.982 &  --- & 1.031 & 1.023 & 1.026 & 1.040 &  --- & 1.033 \\
PTW TM30002/30011 (Farmer) & 0.979 & 0.979 &  --- & 1.035 & 1.026 & 1.030 & 1.043 &  --- & 1.032 \\
PTW TM30004/30012 (Farmer) & 0.972 &  --- &  --- & 1.042 &  --- & 1.037 &  --- &  --- & 1.039 \\
PTW TM30006/30013 (Farmer) & 0.982 & 0.981 &  --- & 1.032 & 1.024 & 1.027 & 1.041 &  --- & 1.036 \\
PTW 31002 flexible & 0.983 & 0.983 &  --- & 1.030 & 1.022 & 1.025 & 1.039 &  --- &  --- \\
PTW 31003 flexible & 0.983 & 0.983 &  --- & 1.030 & 1.022 & 1.025 & 1.039 &  --- &  --- \\
PTW 31006 PinPoint & 0.988 & 0.990 &  --- & 1.025 & 1.015 & 1.020 & 1.032 &  --- &  --- \\
PTW 31014 PinPoint & 0.987 & 0.998 &  --- & 1.026 & 1.007 & 1.021 & 1.023 &  --- &  --- \\
PTW 31016 PinPoint &  --- & 0.999 &  --- &  --- & 1.005 &  --- & 1.022 &  --- &  --- \\

\br
\end{tabular}
\end{indented}
\end{table}

\addtocounter{table}{-1}

\Table{Pertubation and beam quality correction factors for carbon ion beams (\textit{continued}).}
\br

& & $p_{^{60}\rm{Co}}$ & & & & & $k_Q$ & & \\
Chamber type	& \scriptsize{TRS398} & \scriptsize{DIN6801-1} & \scriptsize{Gom\`a \emph{et al.}} & \scriptsize{TRS398} & \scriptsize{DIN6801-1} & & \scriptsize{This study} & & \scriptsize{Experiment} \\
& \tiny{(2006)} & \tiny{(2010)} & \tiny{(2016)} & \tiny{(2006)} & \tiny{(2010)} & \tiny{$p$ TRS398} & \tiny{$p$ DIN} & \tiny{$p$ Gom\`a} & \\
\mr

SNC 100700-0 Farmer & 0.982 &  --- &  --- & 1.031 &  --- & 1.026 &  --- &  --- &  --- \\
SNC 100700-1 Farmer & 0.972 &  --- &  --- & 1.042 &  --- & 1.037 &  --- &  --- &  --- \\
Victoreen Radocon III 550 & 0.983 &  --- &  --- & 1.030 &  --- & 1.025 &  --- &  --- &  --- \\
Victoreen Radocon II 555 & 1.001 &  --- &  --- & 1.012 &  --- & 1.007 &  --- &  --- &  --- \\
Victoreen 30-348 & 0.991 &  --- &  --- & 1.022 &  --- & 1.017 &  --- &  --- &  --- \\
Victoreen 30-351 & 0.989 &  --- &  --- & 1.024 &  --- & 1.019 &  --- &  --- &  --- \\
Victoreen 30-349 & 0.985 &  --- &  --- & 1.028 &  --- & 1.023 &  --- &  --- &  --- \\
Victoreen 30-361 & 0.992 &  --- &  --- & 1.021 &  --- & 1.016 &  --- &  --- &  --- \\
& & & & & & & & & \\
	\textit{Plate-parallel chambers} & & & & & & & & & \\
Attix RMI 449 &  --- &  --- &  --- & 0.990 &  --- & 0.985 &  --- &  --- &  --- \\
Capintec PS-033 &  --- &  --- &  --- & 1.024 &  --- & 1.019 &  --- &  --- &  --- \\
Exradin A10 &  --- & 0.967 & 0.996 &  --- & 1.039 &  --- & 1.056 & 1.025 &  --- \\
Exradin A11 &  --- & 0.985 & 0.983 &  --- & 1.020 &  --- & 1.037 & 1.039 &  --- \\
Exradin A11TW &  --- &  --- & 0.975 &  --- &  --- &  --- &  --- & 1.048 &  --- \\
Exradin P11 &  --- & 1.034 &  --- & 0.995 & 0.971 & 0.990 & 0.988 &  --- &  --- \\
Exradin P11TW &  --- & 1.039 &  --- &  --- & 0.967 &  --- & 0.983 &  --- &  --- \\
Holt (Memorial) &  --- &  --- &  --- & 1.009 &  --- & 1.004 &  --- &  --- &  --- \\
IBA---CP &  --- & 1.020 &  --- & 0.989 & 0.985 & 0.984 & 1.001 &  --- &  --- \\
IBA---CP-02 &  --- &  --- & 1.023 &  --- &  --- &  --- &  --- & 0.998 &  --- \\
IBA PPC-05 &  --- & 1.013 & 1.010 &  --- & 0.992 &  --- & 1.008 & 1.012 & 0.987 \\
IBA PPC-40 &  --- & 1.010 & 1.012 &  --- & 0.995 &  --- & 1.011 & 1.009 & 0.988 \\
PTW TM34045 Adv. Markus &  --- & 1.010 & 1.018 &  --- & 0.995 &  --- & 1.011 & 1.003 &  --- \\
PTW TM23343 Markus &  --- & 1.008 & 1.015 & 1.004 & 0.997 & 0.999 & 1.013 & 1.006 &  --- \\
PTW TM34001 (Roos) &  --- & 1.014 & 1.013 & 1.003 & 0.991 & 0.998 & 1.007 & 1.009 & 0.999 \\
\br
\end{tabular}
\end{indented}
\end{table}

\end{landscape}

\subsection{Error estimation}

Tabs.~\ref{tab:quantities} and \ref{tab:uncertainty} list the standard uncertainty for the stopping power ratios and other input quantities. No major change was introduced by the ICRU90 report. $^{60}$Co-related inaccuracy is still dominated by the uncertainty in perturbation factors as given in the original TRS398 report, especially for plate-parallel chambers. The recalculation of stopping power ratio in this study allows however to reduce the corresponding standard uncertainty. Although the inaccuracy in stopping powers for carbon ions given in ICRU90 is larger for low energies than for protons, most factors in the computation of stopping power for water and air can be assumed to be correlated and should therefore be mitigated by taking the ratio. The major contribution comes then from the uncertainty in I value. A worst case scenario, i.e. the opposite deviation of these values, yields 1.5\,\% standard uncertainty as a conservative upper limit. The statistical error from Monte Carlo simulations in this study is negligible while the estimation of other factors such as the systematic errors from physical models are out of the scope of this study but can be assumed to be negligible as well in the context of stopping power ratio estimation. Using a constant stopping power ratio, the assignment of a beam quality contributes 0.3\,\% uncertainty. In contrast, Hartmann~\emph{et al.}~\cite{Hartmann1999} summarized both effects to 2\,\%. The numbers in Tab.~\ref{tab:uncertainty} underline the necessity for full radiation transport simulations providing chamber specific values $f=s_{\rm{w,air}}\cdot p$ -- for which the additions to the \textsc{Geant4} toolkit from this study are a prerequisit -- and experimental data on $W/e$ for light ions.

\begin{landscape}
\Table{Uncertainty budget for the computation of beam quality correction factor $k_Q$ for carbon ion beams.}
\br
 & Chamber type & cylindrical & plate-parallel & cylindrical & plate-parallel \\
 & Beam & carbon & carbon & carbon+$^{60}$Co & carbon+$^{60}$Co\\
Component & & $u_c$ (\%) & $u_c$ (\%) & $u_c$ (\%) & $u_c$ (\%) \\
\mr
$s_{\rm{w,air}}$ & & 1.5 & 1.5 & 1.6 & 1.6\\
Assingment to beam quality & & 0.3 & 0.3 & 0.4 & 0.4 \\
$W/e$ & & 1.5 & 1.5 & 1.5 & 1.5\\
$p$ & & 1.0 & 1.0 & 1.2 & 1.8 \\
\br
Total & & 2.4 & 2.4 & 2.5 & 2.9 \\
\br
\end{tabular}
\end{indented}
\label{tab:uncertainty}
\end{table}
\end{landscape}

\section{Conclusion}
The new key data on stopping power tables from the ICRU90 Report were implemented in the Geant4 toolkit for water-to-air stopping-power ratio calculations for monoenergetic and SOBP carbon ion beams. Results of $s_{\rm{w,air}}$ for monoergetic carbon ions were shown to agree within 0.1\,\% to previously published calculations providing confidence for the evaluation of SPR in different reference conditions. 
The impact of integration limits as well as the choice of electronic or total stopping power on the stopping-power ratio computation was shown to be negligible.\\

New recommendations for the water-to-air stopping-power ratio are presented, namely, $s_{\rm{w,air}} = 1.1247$ for the reference condition of 1\,g\,cm$^{-2}$ depth for monoenergetic carbon ions, and $s_{\rm{w,air}} = 1.1274$ at the center of physically-optimized SOBPs. Parametrizations of $s_{\rm{w,air}}$ with respect to residual range in water were obtained for the reference conditions of monoenergetic carbon ion beams and SOBPs. These can be applied to precisely estimate $s_{\rm{w,air}}$ at the different reference conditions investigated in this study. Eventually, it was shown that the new recommendation of a constant stopping-power ratio $s_{\rm{w,air}} = 1.126$ represents the variation of $s_{\rm{w,air}}$ for different reference conditions within 0.3\,\% which is considerably smaller than a conservative estimate of the uncertainty connected with SPR data ($1.5\,\%$).\\
 
The impact of the resulting $s_{\rm{w,air}}$ for ions together with the updated key data on the beam quality correction factors for carbon ion beams was evaluated. The changes due to the updated key quantities were found to agree very well with experimental data for the case of cylindrical chambers when using $^{60}$Co perturbation factors from TRS398 or Gom\`a et al~\cite{Goma2016}. An average between these two and experimental data -- where available -- should be used as a recommendation. The beam quality correction factor remains the dominant contributor to the uncertainty in reference dosimetry for light ions. Experimental data from calorimetry is key in reducing this uncertainty in the future -- if possible studying also plane-parallel chambers more intensively.

\section*{Acknowledgements}
The authors are deeply grateful to Prof. Pedro Andreo and Prof. Steve Seltzer for fruitful discussions and making the BEST software available.

\end{document}